 \documentclass[final,1p,times]{elsarticle}

\usepackage{amssymb}
\usepackage{xcolor}
\usepackage[flushleft]{threeparttable}
\usepackage{array,booktabs,makecell}
\usepackage[font=small]{caption}

\journal{Physics Letters B}

\begin{document}

\begin{frontmatter}



\title{$^{57}{\rm Zn}$ $\beta$-delayed proton emission establishes the $^{56}{\rm Ni}$ $rp$-process waiting point bypass}

\author[OU]{M.~Saxena \corref{cor1}}
\ead{saxenam@ohio.edu}
\author[LL]{W.-J~Ong}
\author[OU]{Z.~Meisel}
\author[LL,LW]{D.~E.~M.~Hoff}
\author[FR]{N.~Smirnova}
\author[LW]{P.~C.~Bender}
\author[LL]{S.~P.~Burcher}
\author[AR]{M.~P.~Carpenter}
\author[DV]{J.~J.~Carroll}
\author[NSCL]{A.~Chester}
\author[DV]{C.~J.~Chiara}
\author[OU]{R.~Conaway}
\author[AR]{P.~A.~Copp}
\author[MS]{B.~P.~Crider}
\author[OU]{J.~Derkin}
\author[CMU]{A.~Estrad\'{e}}
\author[OU]{G.~Hamad}
\author[LL]{J.~T.~Harke}
\author[NSCL,MU]{R.~Jain}
\author[AR]{H.~Jayatissa}
\author[NSCL,MUC,JINA]{S.~N.~Liddick}
\author[NSCL,MU]{B.~Longfellow}
\author[NSCL,MUC]{M.~Mogannam}
\author[NSCL]{F.~Montes}
\author[CMU]{N.~Nepal}
\author[NSCL,MS]{T.~H.~Ogunbeku}
\author[NSCL,JINA]{A.~L.~Richard\fnref{fn1}}
\fntext[fn1]{Present Address: Lawrence Livermore National Laboratory, Livermore, California 94550, USA}
\author[NSCL,MU,JINA]{H.~Schatz}
\author[OU]{D.~Soltesz}
\author[OU]{S.~K.~Subedi}
\author[CMU]{I.~Sultana}
\author[OR]{A.~S.~Tamashiro}
\author[FL]{V.~Tripathi}
\author[NSCL,MS]{Y.~Xiao}
\author[OU]{R.~Zink}

 \cortext[cor1]{Corresponding author}

\affiliation[OU]{organization={Institute of Nuclear \& Particle Physics, Department of Physics \& Astronomy},
         addressline={Ohio University}, 
city={Athens},
 postcode={45701}, 
state={Ohio},
country={USA}}

\affiliation[LL]{organization={Lawrence Livermore National Laboratory},
            city={Livermore},
            postcode={94550}, 
            state={California},
            country={USA}}

\affiliation[LW]{organization={Department of Physics and Applied Physics, University of Massachusetts Lowell},
            city={Lowell},
            postcode={01854}, 
            state={Massachusetts},
            country={USA}}   
\affiliation[FR]{organization={CENBG, CNRS/IN2P3 and Universit\'{e} de Bordeaux},
            city={Chemin du Solarium},
            postcode={33175}, 
            state={Gradignan Cedex},
            country={France}}  
            
\affiliation[AR]{organization={Argonne National Laboratory},
            city={Argonne},
            postcode={60439}, 
            state={Illinois},
            country={USA}} 

\affiliation[DV]{organization={DEVCOM/Army Research Laboratory},
            city={Adelphi},
            postcode={20783}, 
            state={Maryland},
            country={USA}} 
\affiliation[NSCL]{organization={National Superconducting Cyclotron Laboratory, Michigan State University},
             city={East Lansing},
            postcode={48824}, 
            state={Michigan},
            country={USA}} 

\affiliation[MS]{organization={Department of Physics \& Astronomy, Mississippi State University},
            city={Starkville},
            postcode={39762}, 
            state={Mississippi},
            country={USA}}  

\affiliation[CMU]{organization={Department of Physics, Central Michigan University},
            city={Mount Pleasant},
            postcode={48859}, 
            state={Michigan},
            country={USA}} 

\affiliation[MU]{organization={Department of Physics \& Astronomy, Michigan State University},
            city={East Lansing},
            postcode={48824}, 
            state={Michigan},
            country={USA}}  
            
\affiliation[MUC]{organization={Department of Chemistry, Michigan State University},
            city={East Lansing},
            postcode={48824}, 
            state={Michigan},
            country={USA}} 

\affiliation[JINA]{organization={Joint Institute for Nuclear Astrophysics -- Center for the Evolution of the Elements, Michigan State University},
            city={East Lansing},
            postcode={48824}, 
            state={Michigan},
            country={USA}}
            
\affiliation[OR]{organization={Department of Physics, Oregon State University},
            city={Corvallis},
            postcode={ 97331}, 
            state={Oregon},
            country={USA}}
            
\affiliation[FL]{organization={Department of Physics, Florida State University},
            city={Tallahassee},
            postcode={32306}, 
            state={Florida},
            country={USA}} 

\begin{abstract}
We  measured the $^{57}{\rm Zn}$ $\beta$-delayed proton ($\beta p$) and $\gamma$ emission at the National Superconducting Cyclotron Laboratory. We find a $^{57}{\rm Zn}$ half-life of $43.6\pm0.2$ ms, $\beta p$ branching ratio of $(84.7\pm1.4)\%$, and identify four transitions corresponding to the exotic $\beta$-$\gamma$-$p$ decay mode, the second such identification in the $fp$-shell. The $p$/$\gamma$ ratio was used to correct for isospin mixing while determining the $^{57}{\rm Zn}$ mass via the isobaric multiplet mass equation. Previously, it was uncertain as to whether the $rp$-process flow could bypass the textbook waiting point $^{56}{\rm Ni}$ for astrophysical conditions relevant to Type-I X-ray bursts. Our results definitively establish the existence of the $^{56}{\rm Ni}$ bypass, with 14-17\% of the $rp$-process flow taking this route.
\end{abstract}

\end{frontmatter}



Type-I X-ray bursts are explosions lasting tens of seconds on the surface of an accreting neutron star, resulting in neutron star envelope temperatures near 1~GK. These bursts are thermonuclear explosions powered by a sequence of nuclear reactions known as the rapid-proton capture ($rp$)-process, which involves hundreds of proton-rich nuclides~\cite{Wallace,parikh,Meis18}. The flow of the $rp$-process nuclear reaction sequence in astrophysics model calculations is sensitive to nuclear physics inputs, and thereby, so are the calculated burst light curve and ashes~\cite{Parikh_2008,Cybu16,Scha17,Meis19}. Of particular interest are nuclear waiting-points~\cite{VanW94,Scha98,Woos04}. At a waiting-point nucleus, a low proton-capture $Q$-value ($Q_{p,\gamma}\lesssim1$~MeV) leads to
$(p,\gamma)-(\gamma,p)$ equilibrium, which stalls nucleosynthesis until a weak decay or proton-capture on the small
equilibrium abundance of the waiting-point $(p,\gamma)$ product finally takes place.

Of the waiting-points, $^{56}{\rm Ni}$ is by far the most influential, leading to the characteristic peak in the X-ray burst light curve~\citep{Wallace,Kan,Lang14}.
However, an alternative pathway circumventing $^{56}{\rm Ni}$  has been proposed~\citep{2001forst,Ong,Val}, where the $rp$-process
flow could instead be diverted through the reaction sequence $^{55}{\rm Ni}(p,\gamma)^{56}{\rm Cu}(p,\gamma)^{57}{\rm Zn}(\beta^{+})
^{57}{\rm Cu}(p,\gamma)^{58}{\rm Zn}$, bypassing  $^{56}{\rm Ni}$ completely. A recent precision mass measurement of $^{56}{\rm Cu}$ constrained the  $^{55}{\rm Ni}(p,\gamma)^{56}{\rm Cu}$ reaction rate to within a factor of two and determined the temperature-density phase space
 where  $^{55}{\rm Ni}$ and  $^{56}{\rm Cu}$ come into (p,$\gamma$)-($\gamma$,p) equilibrium~\citep{Val,Valv2}. The
 $^{56}{\rm Cu}$(p,$\gamma$)$^{57}{\rm Zn}$ reaction  rate is likely uncertain to within an order of magnitude~\cite{Fisk01}.
However, since $Q_{p,\gamma}=1.2\pm0.2$~MeV~\cite{Huan21}, the forward reaction rate always exceeds the reverse, enabling nucleosynthesis to flow to $^{57}{\rm Zn}$. As such, the dominant uncertainty for the
$^{56}{\rm Ni}$ bypass is the probability that $^{57}{\rm Zn}$ undergoes $\beta$-delayed proton ($\beta p$) emission, which
would re-populate $^{56}{\rm Ni}$.

The $^{57}{\rm Zn}$ $\beta p$ branching ratio is presently constrained to $(90\pm10)\%$~\cite{Ciem20}. Prior measurements found a $(78\pm17)\%$  $\beta p$ branching~\cite{2007blank}. Both, these results are consistent with either completely blocking the $^{56}{\rm Ni}$ bypass or, for a 1$\sigma$ lower bound, 
up to 20\% of the $rp$-process flow in the N~=~27 isotonic chain following the bypass over the range of envelope conditions relevant for X-ray bursts. Therefore, a higher-precision $^{57}{\rm Zn}$ $\beta p$ branching ratio is required for a meaningful constraint on the $^{56}{\rm Ni}$ bypass.

Meanwhile, $\beta p$ emission is also a valuable probe of isospin-mixing in exotic nuclides. The exotic $\beta$-$\gamma$-$p$ decay mode has recently been observed in $^{56}{\rm Zn}$ (in the $fp$-shell), including $\gamma$ decays from the isobaric analogue state (IAS)~\cite{Orri14,Orri16}. The $p/\gamma$ decay ratio can be used to determine isospin mixing for the IAS~\cite{Smir16,Smir17}. Taking this isospin mixing into account for the IAS is essential for obtaining accurate nuclear mass constraints from the isobaric multiplet mass equation (IMME)~\cite{Macc14,Su16}. For instance, constraints on isospin mixing of the $^{57}{\rm Cu}$ IAS are required for an IMME mass determination of $^{57}{\rm Zn}$ and, consequently, of the $^{56}{\rm Cu}(p,\gamma)^{57}{\rm Zn}$ $Q_{p,\gamma}$ value.

We report results from a $\beta p$ emission measurement of $^{57}{\rm Zn}$ performed at the National Superconducting Cyclotron Laboratory (NSCL), that substantially reduce the main nuclear physics uncertainties associated with the $^{56}{\rm Ni}$ bypass in the $rp$-process and identify the second case of $\beta$-$\gamma$-$p$ decay in the $fp$-shell. 

\begin{figure}[ht!]
\begin{center}
\includegraphics[width=1.0\columnwidth]{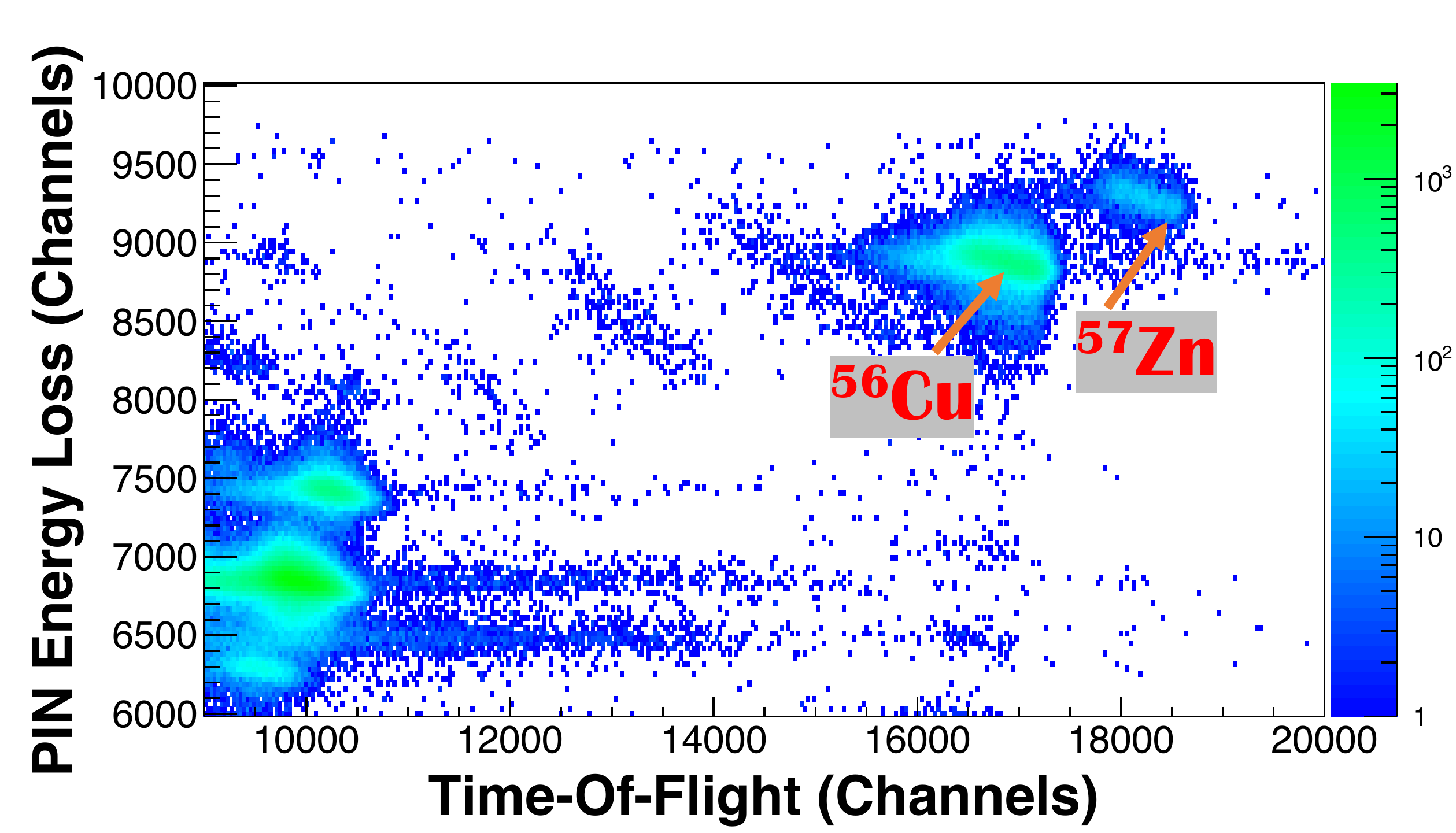}
\end{center}
\caption{(color online) Particle identification (PID) plot used to identify $^{57}{\rm Zn}$. Time-of-flight between the A1900 extended focal plane and the Si PIN detector is plotted on the horizontal axis while energy loss in the Si PIN detector is plotted on the vertical axis.
}
\label{fig:pid}
\end{figure}

In this experiment, $^{57}{\rm Zn}$ was produced by fragmentation
of a 150 MeV/nucleon $^{78}{\rm Kr}$ primary beam on a 305 mg/cm$^2$  beryllium target
at the Coupled Cyclotron Facility of NSCL. The
A1900 fragment separator~\citep{Morr03} was used to perform the initial beam purification
via the  $B\rho$-$\Delta$E-$B\rho$ technique, 
using a 450 mg/cm$^2$ Al wedge at the
intermediate image position and $\pm$0.5$\%$ momentum acceptance. We used the Radio Frequency Fragment Separator~\cite{Bazi09} to remove contamination from the low-momentum tail of nuclides closer to stability. Particle identification was performed event-by-event via $\Delta E-$TOF, using energy loss from a Si PIN detector upstream of the ion implantation station, and time-of-flight between the Si PIN detector and a scintillator in the A1900 extended focal plane, see Fig.~\ref{fig:pid}

13860 $^{57}{\rm Zn}$ ions were implanted into a $525$-$\mu$m thick double-sided silicon strip detector (DSSD) segmented into 40$\times$40 1-mm pitch strips located within the Beta Counting Station (BCS)~\cite{Pris03}. A 998-$\mu$m thick single-sided silicon
detector located downstream of the DSSD was used as a light-ion veto.
The BCS was surrounded by the Clovershare array, consisting of 16 high
purity germanium clover detectors, for $\gamma$ detection.
A front-back coincidence was required for a valid  DSSD signal,  where
implants and $\beta$-delayed protons were time-stamped, allowing for implant-decay correlations. The spatial and time correlation window used in the analysis were a $3\times3$ pixel area and $t_{\rm corr}$ set equal to 3000~ms, respectively. The long correlation window was used to allow the estimation of the background decays from 1000--3000~ms. 

\begin{figure}[ht!]
\begin{center}
\includegraphics[width=1.0\columnwidth]{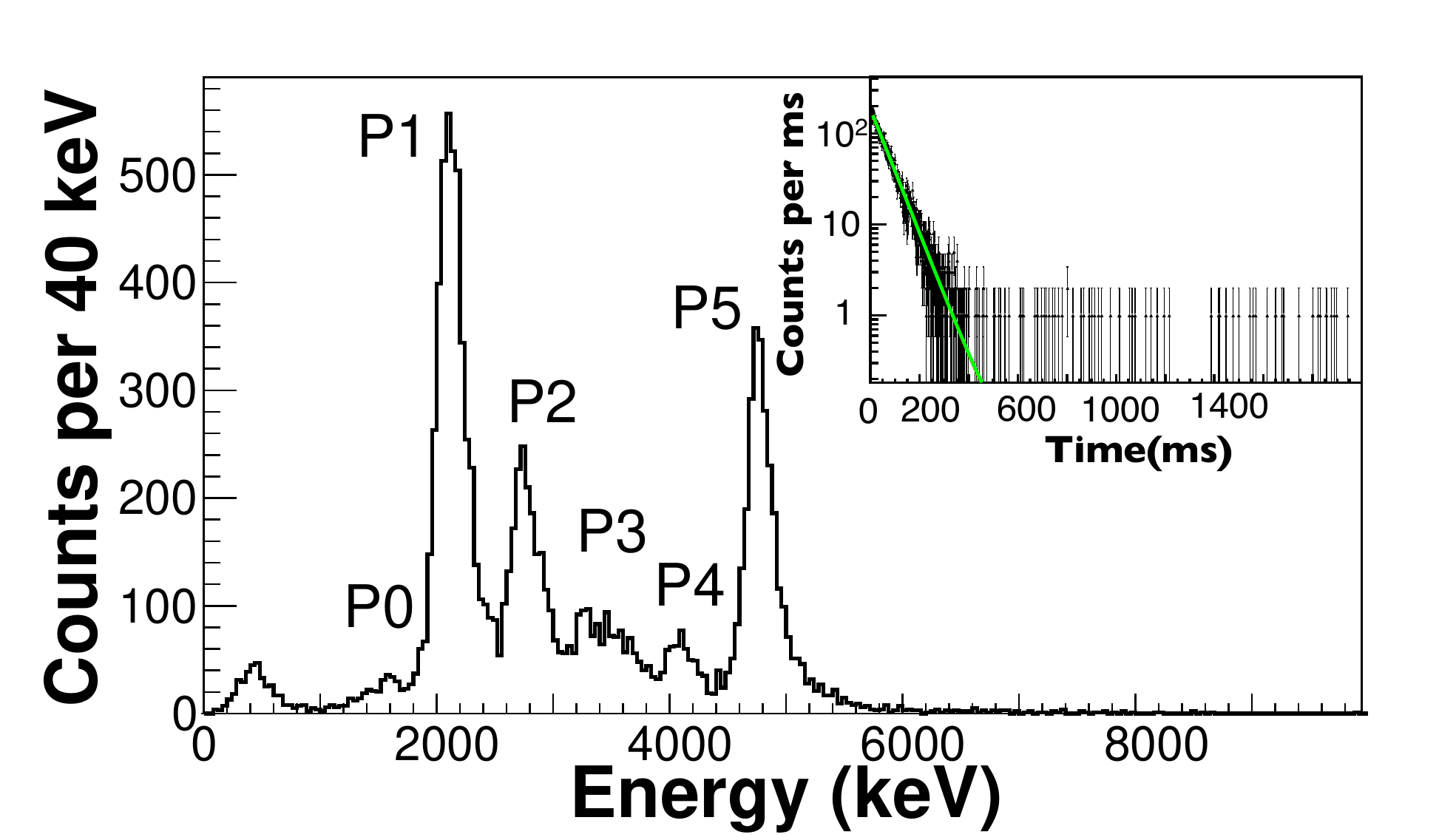}
\end{center}
\caption{(color online) Particle decay energy spectrum measured in the DSSD for the decay events correlated with $^{57}{\rm Zn}$ implants, with the observed proton groups labeled from P0--P5. The inset shows the decay curve of the correlated $^{57}{\rm Zn}$ events. The green line indicates the best fit results when employing an exponential decay plus a constant background for events with $E_{\rm dec}~>$~900 keV.}
\label{fig:dssd}
\end{figure}

A $^{57}{\rm Zn}$ nucleus can $\beta$ decay by either emitting a $\beta ^{+}$ and/or a   $\beta ^{+}$ followed by a proton. The protons and the $\beta ^{+}$ particle are emitted almost simultaneously from the site of the implantation in the DSSD. Since the $\beta$'s usually escape the implantation detector by depositing only a fraction of their energy; pure $\beta$ decays contribute to mostly the low-energy spectrum. Protons on the other hand are stopped inside the DSSD and deposit their full energy. Events below 900~keV were associated with $\beta$-only decays. This assumption was made by overlaying the $scaled$ decay energy spectrum of $^{56}$Cu $\beta$ decay, which was also implanted in the DSSD. For every 100 $^{56}$Cu $\beta$ decays, only  $0.4\%$ have the $\beta$-delayed proton branch~\cite{Borc01}. The particle decay energy spectrum obtained in the present analysis for $^{57}{\rm Zn}$ decays is shown in Fig.~\ref{fig:dssd}, with decay ($\beta+p$) energies $E_{\rm dec}$ ranging from $900-6000$~keV.
The inset of Fig.~\ref{fig:dssd} shows the decay curve, including an exponential fit and fixed background, which is gated on decays with $E_{\rm dec}$ $>$ 900~keV, corresponding to a $^{57}{\rm Zn}$ half-life $t_{1/2}$ of $43.6\pm0.2$~ms.
This is in satisfactory agreement with the prior determinations of $40\pm10$~ms in Ref.~\cite{Viei76} and $48\pm3$~ms in Ref.~\cite{2007blank}, but in tension with the Ref.~\cite{Ciem20} result of $27\pm3$~ms,  which reported 25 times lower statistics in comparison to our measurement.

Peaks in Fig.~\ref{fig:dssd} were fitted using a convolution
of a Gaussian distribution defining the proton emission with a Landau distribution for the high-energy tail caused
by $\beta$-summing~\cite{meisel17}. The width of the Gaussian distribution was fixed by the intrinsic resolution ($\sigma \sim$ 64 keV for 6.5 MeV alpha source from a Thorium-228 radioactive source) of the detector, while the width of the Landau distribution was treated as a free parameter. The centroids of the peaks P0--P5 need to be corrected for $\beta$-summing in order to match the peak energies of the protons reported in Ref.~\cite{jok02}. Henceforth, all the peak centroids in our decay energy spectrum are shifted up by $\sim$ 160 keV.
The resulting delayed-proton intensities are consistent with the results from Ref.~\cite{jok02}; however, they are less precise in the current work due to significant $\beta$-summing and limited resolution of the DSSD.  As a result, the labeled peaks might contain multiple proton transitions as measured in Ref.~\cite{jok02} that could not be resolved in the present work. The proton detection efficiency is inferred to be 100\%. This is based on the consistency of the ratio of counts within decay peaks compared to the results of Ref.~\cite{jok02}. Using peak labels from Fig.~\ref{fig:dssd}, we find P5/P1=$0.59\pm0.03$,  compared to $0.62\pm0.01$ from Ref.~\cite{jok02}, where we have applied our experimental resolution to their data to ensure a consistent comparison. Given the prediction from LISE++~\cite{Tara08} that all implants are within a 100-$\mu$m depth range, it is unlikely such agreement could be obtained if the implantation depth distribution were not roughly centered within the implantation DSSD. A roughly central mean implantation depth is consistent with the minimal asymmetry observed for the decay spectrum peaks~\cite{meisel17}. GEANT4 simulations indicate that for this scenario all protons would be stopped within the implantation DSSD, which is a sufficient but not necessary condition for all $\beta$-delayed proton events to be above the detection threshold.

The total
$\beta$-delayed proton branching ratio $P_{\beta p }$ is the ratio of $\beta$-delayed proton decays ($N_{\beta p}$) to implants ($N_{\rm imp}$). $N_{\beta p}$ can be defined as 
\begin{equation}
N_{\beta p} = \frac {A_{0}}{\lambda} \left(1-e^{-\lambda t_{\rm corr}}\right) \frac{1}{f_{\rm live}}
\end{equation}
and was determined using the proton-gated decay curve in the inset of Fig.~\ref{fig:dssd}, in particular, the initial proton activity $A_{0}$ and decay constant $\lambda$ of the parent $^{57}{\rm Zn}$, including the decay correlation time $t_{\rm corr}$ and correcting for the data acquisition live time $f_{live}$. The uncertainty for $P_{\beta p}$ was calculated through standard error propagation on the above quantities. We find $P_{\beta p}=(84.7\pm1.4$)\% for $^{57}{\rm Zn}$, which is far more precise than but in agreement with $(78\pm17)\%$ from Ref.~\cite{2007blank} and $(90\pm10)\%$ from Ref.~\cite{Ciem20}.

\begin{figure}[ht!]
\begin{center}
\includegraphics[width=1.0\columnwidth]{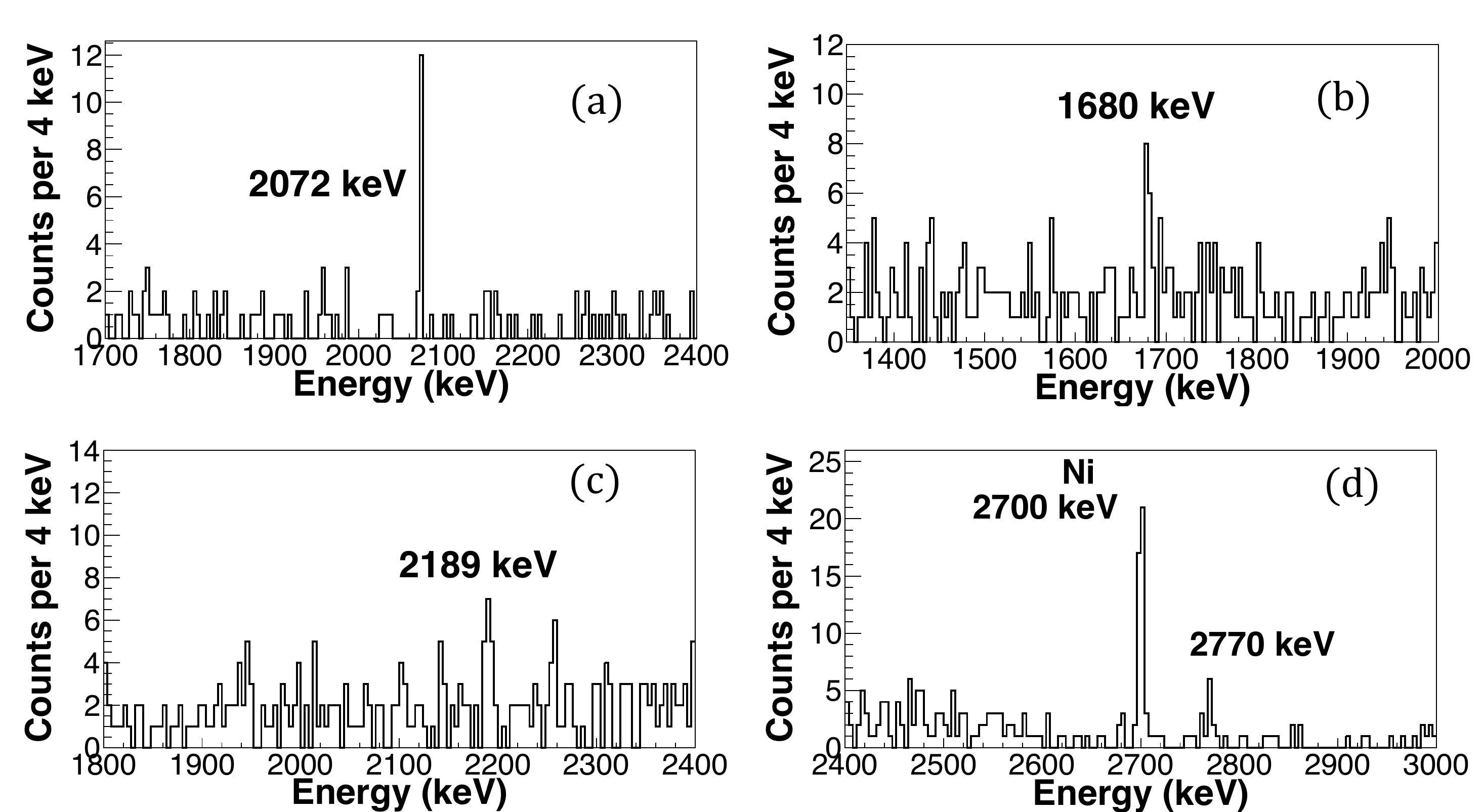}
\caption{The implant-decay correlated $\gamma$-ray spectrum in coincidence with the proton peaks P2 at $E_{\rm dec}$$\sim$$2.7$~MeV (a) and P1 at $\sim2.2$~MeV (b, c, d). Four new $\gamma$-ray transitions with energies 2072~keV, 1680~keV, 2189~keV  and 2770~keV  are observed which constitute the uncommon $\beta$-$\gamma$-p decay mode that occurs due to the small proton separation energy (690~keV) in $^{57}{\rm Cu}$. The 2700~keV $\gamma$-ray transition follows from the de-excitation of the first 2$^{+}$ excited state to the ground state of $^{56}{\rm Ni}$. For completeness, no $\gamma$ rays were observed in coincidence with the proton peaks P3, P4 and P5.} 
\label{fig:gamma}
\end{center}
\end{figure}

Proton-gated $\gamma$-ray coincidence spectra, shown in Fig.~\ref{fig:gamma}, were obtained for the first time for the
peaks labeled in the decay energy spectrum using the Clovershare Array. 
Gating on the proton peak P2 with $E_{\rm dec}$$\sim$$2.7$~MeV  results in a 2072$\pm$0.3~keV $\gamma$-ray transition, which corresponds to the de-excitation
of the $^{57}{\rm Cu}$ IAS ($T$~=~3/2, $J^{\pi}$~=~7/2$^{-}$) at 5297~keV to the proton unbound state at 3225~keV. The proton peak P1 with $E_{\rm dec}$$\sim$$2.2$~MeV 
is coincident with $\gamma$-ray lines at 2770$\pm$0.3~keV and 1680$\pm$0.4~keV, where the former corresponds to de-excitation from the
$^{57}{\rm Cu}$ IAS to the 2530~keV excited state and the latter is consistent with $\gamma$ decay between the 4208~keV
and 2530~keV states in $^{57}{\rm Cu}$. The 2189$\pm$0.8~keV $\gamma$-ray transition is also coincident with the proton peak P1 and is tentatively placed feeding the 2530~keV excited state in $^{57}{\rm Cu}$.
All of these observed $\gamma$-ray transitions are feeding excited states of $^{57}{\rm Cu}$ that are proton unbound and belong to an exotic $\beta$-$\gamma$-$p$ decay mode.  Note that the 2700~keV $\gamma$-ray line corresponds to de-excitation from the first 2$^{+}$ excited state of $^{56}{\rm Ni}$. The level energies in $^{57}{\rm Cu}$ are taken directly from Ref~\cite{jok02}.

\begin{figure*}[ht!]
\begin{center}
\includegraphics[width=1\textwidth,height=10.0cm]{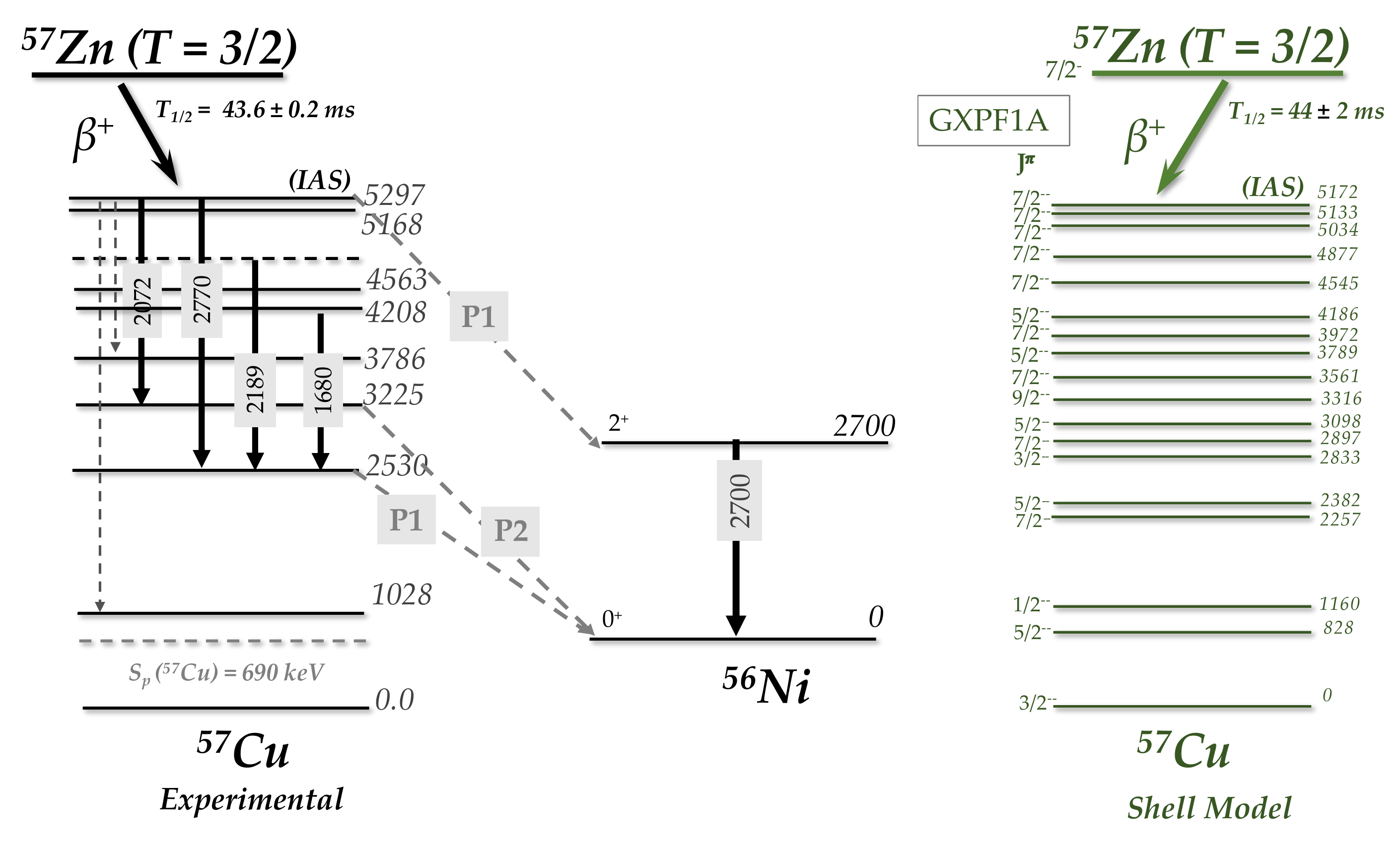}
\caption{The new partial decay scheme of $^{57}{\rm Zn}$~$\beta$p and $\beta$-$\gamma$-p decay built  using the proton and $\gamma$-ray coincidences. The  theoretical partial level scheme from shell model calculation using GXPF1A interaction is shown on the right. The dashed  $\gamma$-ray lines correspond to transitions observed in the mirror nucleus $^{57}{\rm Ni}$ but not detected in our present experiment. The 2189~keV $\gamma$-ray line was placed according to a similar $\gamma$-ray line seen in the mirror nucleus $^{57}{\rm Ni}$, though it may not be unambiguous. The intensity of 2189~keV accounts for the missing  $I_{\beta}$ strength. } 
\label{fig:scheme}
\end{center}
\end{figure*}

The new partial decay scheme for $^{57}{\rm Zn}$ resulting from this work, including the four
newly observed $\beta$-$\gamma$-$p$ branches, is shown in Fig.~\ref{fig:scheme}. It is noteworthy to restate that the energies of the labeled $\gamma$'s in Fig.~\ref{fig:scheme}, are consistent with the energy differences of the excited states of $^{57}{\rm Cu}$ in the known level scheme~\cite{jok02}.
The Feldman-Cousins statistical approach~\cite{FC} was used to confirm the statistical significance of the $\gamma$-ray peaks.
The relative intensities and the energies of the proton peaks are known from the previous $\beta$-decay measurement by Ref.~\cite{jok02}
which used a $\Delta E$-$E$ (silicon + gas) telescope detector to detect the protons and the positrons.  The absolute intensities of the $^{57}{\rm Cu}$ 
individual proton transitions were determined by renormalizing these previously known relative intensities using our results. The absolute $\beta$-branch intensities were determined by correcting the absolute proton
intensities by the respective intensities of the $\gamma$-ray transitions de-exciting or feeding a state in  $^{57}{\rm Cu}$. An absolute $\gamma$-ray efficiency curve was produced using two sources -- a standard $\gamma$-ray source (SRM - a mixed radionuclide point source consisting of $^{125}{\rm Sb}$, $^{154}{\rm Eu}$ and $^{155}{\rm Eu}$)  for energies from 40 keV up to 1.5 MeV and a $^{56}{\rm Co}$ source for energies up to 3.5 MeV. A flat uncertainty of $\sim 3\%$ was used for all energies, which included the error due to the calibrated activity of the source.
The final experimental $\beta$-intensities~($I_{\beta}$) are noted in Table.~\ref{tab:intensity} along with the $\log(ft)$ values for each level.
The statistical rate function $f$ was determined following the $^{57}{\rm Zn}-^{57}{\rm Cu}$ ground-state to ground-state $Q$ value, 14474(60)~keV, calculated using the present ${\rm ME}(^{57}{\rm Zn})$ (described later in text). The present experimental $\beta$ intensities allow for an estimation of the partial half-lives of the states in the daughter nucleus equal to $T_{1/2}$/$I_{\beta}$ and experimental $\log(ft)$ values. 
The $\log(ft)$ for the 5297~keV IAS in $^{57}{\rm Cu}$ was calculated to be 3.37(5), verifying that this is a super-allowed $\beta$ decay~\cite{2007blank}.

$\beta$-transition intensities are directly related to nuclear matrix elements, the Fermi strengths $B$(F), and Gamow-Teller strengths $B$(GT)~\cite{Orri16}. When isospin is a valid symmetry, fragmentation of the Fermi transition is prevented and the strength is concentrated in a single transition to the IAS. This would limit   $B$(F)$\sim\mid N-Z\mid$ to be equal to 3 for the $^{57}{\rm Zn}$ $\beta$ decay. However, the experimental value for the Fermi strength of the 5297~keV IAS is  $B$(F) = $2.62\pm0.2$, indicating a fragmented IAS. Additional confirmation that the $^{57}{\rm Cu}$ IAS is fragmented is the observation of  $\beta$-delayed $\gamma$-emission together with the isospin-forbidden proton emission from the IAS. The missing Fermi strength is thought to be distributed over other $7/2^{-}$ excited states lying within $\sim$500~keV of the IAS, where previous observations of IAS fragmentation in this mass region have been attributed to isospin mixing~\cite{Orri14,Macc14}.

\begin{table}[htb!]
  \caption{Level energies in $^{57}{\rm Cu}$ and newly observed $\gamma$-ray energies ($E_{\gamma}$) following the $^{57}{\rm Zn}$ $\beta$ decay. The experimentally calculated absolute $\beta$-branch intensities ($I_{\beta} \%$) and  log $ft$ for each level is also shown.}
   \label{tab:intensity}
  \centering 
  \begin{threeparttable}
    \begin{tabular}{cccc}
    \midrule
   $E(^{57}{\rm Cu})^{*}$ keV  & $E_{\gamma}$ keV & $I_{\beta}\%$  & log $ft$ \\
   ~\cite{jok02}  & $Exp$  & $Exp$ & $Exp$ \\
    \midrule
     5297 & $2072\pm0.3$   &  $49.0\pm3.6$   &  3.37(5)       \\ 
          & $2770\pm0.3$   &                 & \\
     5168 &                &   $1.6\pm0.7$   &  4.89(27)        \\
    $4719$ & $2189\pm0.8$   &     &         \\
     4563 &                &  $4.5\pm1.0$    &    4.58(14)      \\
     4208 & $1680\pm0.4$   &  $5.9\pm1.2$    &    4.55(21)      \\
     3786 &                &  $5.8\pm1.2$    &    4.65(12)      \\
     3225 &                & $11.1\pm2.3$    &    4.48(13)      \\
     2530 &                &  $3.2\pm2.0$    &   5.16(42)      \\
     \midrule\midrule
    
    \end{tabular}
    \end{threeparttable}
  \end{table}
  
 We used a two-state mixing model, adding an isospin non-conserving part  to the nuclear Hamiltonian~\cite{Orri14,Smir16}. This  causes the analogue state at 5297(34)~keV ($T_{A}$~=~3/2) and the state closest in energy at 5168(35) keV ($T_{{B}}$~=~1/2) to mix, making the Fermi transition split between these two states. 
The ratio of the Fermi transition strength $B$(F)$_{T_{B}}$ = 0.087(33)/$B$(F)$_{T_{A}}$~=~2.62(20) gives an estimate of the isospin impurity ($\alpha^{2}$) in the $T_{B}$~=~1/2 5168 keV state to be equal to 3(1)\%. Using the observed energy difference of the states, ($E_{A}$-$E_{B}$) keV, the mixing matrix element of 23(5) keV is found.

To get a theoretical description of the $^{57}{\rm Zn}$  $\beta$-delayed decay, we have performed large scale shell-model calculations 
in the full pf-shell model space with the GXPF1A Hamiltonian~\cite{Hon04,Honm05} using the NuShellX@MSU code~\cite{NushellX}. 
The calculated low-energy spectrum of $^{57}{\rm Cu}$  is found to be in good agreement with experiment, see Fig.~\ref{fig:scheme}. The 7/2$^{-}$  IAS is obtained to be the ($n$~=~11) 7/2$^{-}$ state at 5.17 MeV.
Within the experimental $Q$-value window for the $\beta$ decay, including 20 states of each  7/2$^{-}$, 5/2$^{-}$ and 9/2$^{-}$ spin and parity, and applying a quenching factor of
$q$~=~0.77~\cite{Cau94} to the Gamow-Teller (GT) operator, we find $^{57}{\rm Zn}$~$t_{1/2}$~=~44(2) ms, which is in excellent agreement with the experimental value. 
The shell-model calculations confirm that the $B$(GT) strengths for the IAS and nearby 7/2$^{-}$ states are small.
We have also deduced proton and electromagnetic widths of excited states in $^{57}{\rm Cu}$, which will be presented and analyzed in detail 
in a forthcoming publication.

The ground state of $^{57}{\rm Zn}$ ($T_{Z}$~=~$-$3/2) and its isobaric analogue state in the daughter $^{57}{\rm Cu}$ ($T_{Z}$~=~$-$1/2) are part
of an isobaric multiplet of four states ($A$~=~57, $T$~=~3/2). By measuring the energy of the IAS in  $^{57}{\rm Cu}$ as 5301 $\pm$ 34 keV, the IMME was
used to provide an indirect measurement of the mass of the  $^{57}{\rm Zn}$  ground state, taking into account the correction for isospin mixing~\cite{Macc14,Su16}. The other two members of the multiplet are the ground state of $^{57}{\rm Co}$, with an atomic mass excess ${\rm ME}(^{57}{\rm Co})$ of $-59345.6\pm0.5$~keV, and the IAS of $^{57}{\rm Ni}$ at 5239~keV excitation energy, with a corresponding ${\rm ME}(^{57}{\rm Ni})$ of $-50844.8\pm0.5$~keV ~\cite{Wang17,Bhat98}. This results in an IMME-based ${\rm ME}(^{57}{\rm Zn})$ of $-32835\pm60$~keV. Our ${\rm ME}(^{57}{\rm Zn})$ is in agreement with, but more precise, than the 2020 Atomic Mass Evaluation~\cite{Huan21} extrapolation $-32550\pm200$~keV and results based on local mass relations, $-32808\pm86$~keV from Ref.~\cite{Ma} and $-32898\pm70$~keV from Ref.~\cite{zong}. Our result is also congruent with the value of $-32830\pm50$~keV presented in~\cite{2007blank}.

Using our ${\rm ME}(^{57}{\rm Zn})$ and ${\rm ME}(^{56}{\rm Cu})$ from Ref.~\cite{Huan21}, we find 1494(61) keV as the $^{56}{\rm Cu}(p,\gamma$)$^{57}{\rm Zn}$ $Q_{p,\gamma}$~-value. 
We calculated the astrophysical reaction rate using the narrow resonance formalism, where $N_{A}\langle\sigma v\rangle\propto\Sigma_{i}\omega\gamma_{i}\exp(-E_{r,i}/(k_{\rm B}T_{9}))$. Here $T_{9}$ is the temperature in GK, $k_{\rm B}$ is Boltzmann's constant, and a resonance energy is $E_{r,i}=E^{*}(^{57}{\rm Zn})-Q_{p,\gamma}$, calculated from the $^{57}{\rm Zn}$ excitation energy of state $i$ and the $Q_{p,\gamma}$. The resonance strength of state $i$ is 
\begin{equation}
\omega\gamma_{i}=\frac{2J_{r,i}+1}{(2J_{p}+1)(2J_{\rm Cu,g.s.}+1)}\frac{\Gamma_{p}\Gamma_{\gamma}}{\Gamma_{p}+\Gamma_{\gamma}},
\end{equation}
where we adopted resonance spins $J_{r,i}$ and partial $\gamma$-decay widths $\Gamma_{\gamma}$ from the shell-model calculations of Ref.~\cite{Fisk01}, while scaling their proton-decay partial widths $\Gamma_{p}$ using the updated $Q_{p,\gamma}$ (see e.g.~Ref.~\cite{merz}).

\begin{figure}[ht!]
\begin{center}
\includegraphics[width=1.0\columnwidth]{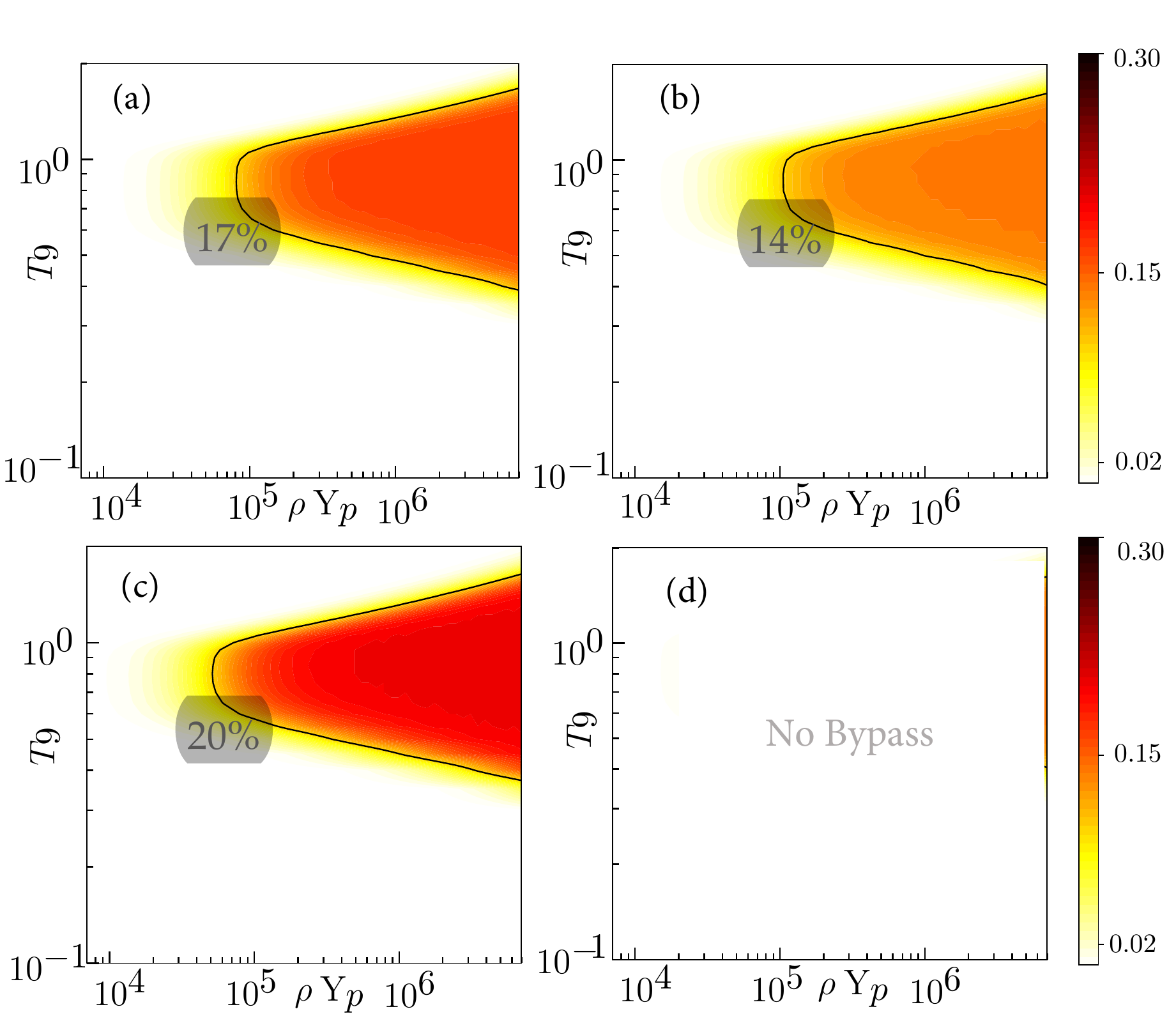}
\caption{ (color online)
$T_{9}-\rho$ phase space showing the  $^{56}{\rm Ni}$ bypass fraction in the $rp$-process, where $T_{9}$ is the temperature in GK and $\rho Y_{p}$ is in units of mol cm$^{-3}$. The present $\pm$1$\sigma$ uncertainties for  $^{57}{\rm Zn}$ $t_{1/2}$, $P_{\beta p}$ and ${\rm ME}(^{57}{\rm Zn})$  result in a maximal  $^{56}{\rm Ni}$ bypass fraction between  17\% and 14\% (a,b) as compared to the previous constraints~\cite{Ciem20} showing a broader bypass range of 20\% to 0\% (c,d).} \label{fig:flow}
\end{center}
\end{figure}

 To study the impact of our $^{57}{\rm Zn}$ $t_{1/2}$, $P_{\beta p}$, and ${\rm ME}(^{57}{\rm Zn})$ on the extent of the $^{56}{\rm Ni}$ bypass in the $rp$-process, we performed flow calculations with a reaction network limited to $N=27-29$ and $Z=28-30$, using our data as discussed and the REACLIBv2.2 library~\cite{Cybu10} otherwise. Calculations were seeded at $^{55}{\rm Ni}$ and run at constant $T_{9}$ and constant density $\rho$ for 1~s.
The flow of the bypass around $^{56}{\rm Ni}$ waiting point  is described as the amount of  $^{57}{\rm Cu}$ isotope $not$ produced by the traditional route of proton capture on $^{56}{\rm Ni}$. This can be defined as the ratio of the flow through $^{57}{\rm Zn}$ $\beta$-only decay to the combined flow through all $\beta$-decay branches in the network. The two extreme cases of maximal and minimal bypass are generated taking 1-$\sigma$ uncertainty in the $^{57}{\rm Zn}$ $t_{1/2}$, $P_{\beta p}$ and ${\rm ME}(^{57}{\rm Zn})$. The maximal (minimal) bypass is seen for 1-$\sigma$ lower (upper) value in the $^{57}{\rm Zn}$  $t_{1/2}$, $P_{\beta p}$ and ${\rm ME}(^{57}{\rm Zn})$.  
The phase space diagrams in Fig.~\ref{fig:flow}  show the region where the $^{56}{\rm Ni}$ bypass will be effective, with the colors and contours indicating the strength of the bypass. In comparison with the results of Ref.~\cite{Ciem20}, we restrict the $^{56}{\rm Ni}$ bypass to within 14--17\%, as compared to the prior range 0--20\%, and establish the existence of the $^{56}{\rm Ni}$ bypass.

We note that a more precise value of ${\rm ME}(^{57}{\rm Zn})$ and spectroscopy of $^{57}{\rm Zn}$ for an improved $^{56}{\rm Cu}(p,\gamma)$ reaction rate will only refine the boundaries of the bypass phase space. 

In summary, we measured $\beta$-delayed proton-emission of $^{57}{\rm Zn}$ at the National Superconducting Cyclotron Laboratory using implantation in a double-sided silicon strip detector surrounded by a germanium clover array for $\gamma$-coincidences. We substantially improved the precision for the proton-emission branching ratio, establishing the bypass of the $^{56}{\rm Ni}$ waiting-point in the $rp$-process. We also identify the second case of  $\beta$-$\gamma$-$p$ decays in the $fp$-shell. A detailed analysis of the $^{57}{\rm Zn}$ decay scheme will be treated in a follow-up work. Further refinement of the $^{56}{\rm Ni}$ bypass will require a mass measurement and spectroscopy of $^{57}{\rm Zn}$.

\section*{Acknowledgments}

This work was funded by the U.S. Department of Energy through Grants No. DE-FG02-88ER40387, DE-SC0019042, DE-NA0003909 and
DE-SC0020451; the U.S. National Science Foundation through Grants No. PHY 1848177~(CAREER), PHY-1430152 (Joint Institute for Nuclear Astrophysics -- Center for the Evolution of the Elements) and  PHY-1913554; the U.S. National Nuclear Security Administration through Grant No. DOE-DE-NA0003906, Nuclear Science and Security Consortium, under Award No. DE-NA0003180 and Argonne National Laboratory under the contract number DE-AC02-06CH11357. A part of this work was performed under the auspices of the U.S. Department of Energy by Lawrence Livermore National Laboratory under Contract DE-AC52-07NA27344.
 This work was partially supported by the IN2P3/CNRS, France, under ENFIA project.
Shell-model calculations have been performed at M\'esocentre de Calcul Intensif Aquitain (MCIA), University of Bordeaux.

 \bibliographystyle{elsarticle-num}
 \bibliography{References}

\end{document}